\documentclass[aps, prl, reprint, showpacs, twocolumn, 10pt, groupedaddress]{revtex4-1}

\usepackage{graphicx, amsmath, dsfont, dcolumn, bm, times, color, braket, amssymb, multirow}

\begin{document}

\bibliographystyle{apsrev4-2}

\title{Andreev bound states in junctions formed by conventional and $\mathcal{PT}$-symmetric non-Hermitian superconductors}

\author{Viktoriia Kornich}
\affiliation{Institut f\"ur Theoretische Physik and Astrophysik, Universit\"at W\"urzburg, 97074 W\"urzburg, Germany }
\author{Bj\"orn Trauzettel}
\affiliation{Institut f\"ur Theoretische Physik and Astrophysik, Universit\"at W\"urzburg, 97074 W\"urzburg, Germany }

\date{\today}

\begin{abstract}
We study theoretically a junction of a $\mathcal{PT}$-symmetric non-Hermitian superconductor (PTS) placed between two conventional superconductors. We show that due to non-Hermitian electron-electron interaction in the PTS region and the combination of symmetries, only discrete values of phases of the conventional superconductors yield solutions for Andreev bound states. Remarkably, in the case of $0$ and $\pi$, we obtain growing and decaying in time Andreev bound states. For $\pi/2$ and $3\pi/2$, there is a Majorana zero mode penetrating through the junction in only one direction forming a quasiparticle supercurrent.  
\end{abstract}

\maketitle

\let\oldvec\vec
\renewcommand{\vec}[1]{\ensuremath{\boldsymbol{#1}}}
{\it Introduction.--} 
Hybrid superconducting structures containing normal parts, such as semiconducting nanowires or insulating layers, are platforms for numerous theoretical and experimental investigations often related to the study of Andreev bound states \cite{meng:prb09, lutchyn:prl10,oreg:prl10,hays:naturephys20,petrashov:prl05, melin:prb22}. The high interest to this field partially follows from the fact that Andreev bound states are largely controlled by the phases of superconductors. Recently, the topic of Andreev molecules has arisen because of their potential use in metrology, quantum information, and quantum manipulations \cite{scherubl:bjn19, pillet:nanolett19, kornich:prr19}.

Unconventional and exotic superconductors yield new physical phenomena and states, e.g. high-T$_c$ superconductors \cite{drozdov:nature19}, spin-triplet superconductors \cite{ran:science19}, heavy-fermion superconductors \cite{moeckli:prb21}, flat-band superconductivity \cite{peltonen:jpcm20}, and superconducting Leggett modes \cite{wan:arxiv21}. Taking into account that the obtained data stemming from certain superconducting materials often lacks a clear theoretical explanation, it is desirable to be able to induce or enhance particular types of superconductivity dynamically, e.g. via light or strain \cite{buzzi:prl21,curtis:prr22,kornich:scipost22}. In these cases, the obtained superconductivity is a non-equilibrium phenomenon. Non-equilibrium dynamics can be considered in frames of different formalisms, e.g. Keldysh Green's functions. One of the most recent related theoretical approaches are non-Hermitian systems, where non-Hermiticity is likely to be a consequence of external influence such as a connection to an external bath. 

Non-Hermitian superconductors are not well-understood yet. They are mainly studied in relation to Majorana zero modes \cite{kawabata:prx19, zhou:prb20, wang:prb21, jing:jpcm22}. However, there are particular properties that they exhibit, e.g. odd-frequency pairing \cite{cayao:prb22} or a distinct response in angle-resolved photoelectron fluctuation spectroscopy \cite{kornich:prr22}. $\mathcal{PT}$-symmetric (parity-time-symmetric) non-Hermitian superconductors are a separate class in this field. $\mathcal{PT}$-symmetry allows for real or complex conjugate energy eigenvalues of non-Hermitian Hamiltonians \cite{bender:prl98, bender:rpp07,mostafazadeh:jmp02, klett:pra17, menke:prb17, kawabata:prb18}. $\mathcal{PT}$-symmetric systems are of high interest in optics due to a vast amount of possible applications \cite{ozdemir:natmater19}.

\begin{figure}[tb]
	\begin{center}
		\includegraphics[width=\linewidth]{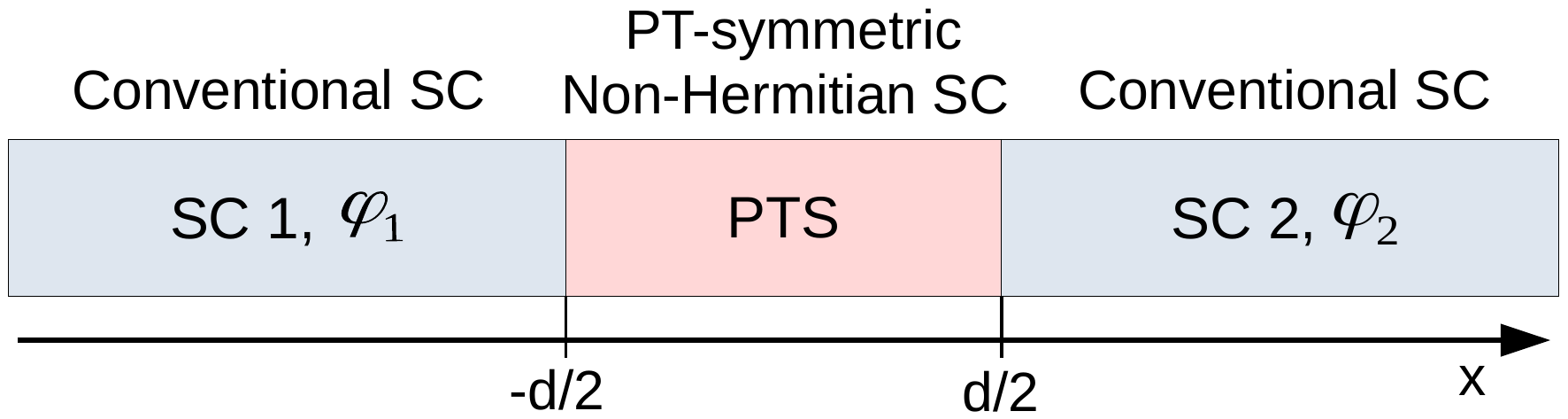}
		\caption{Setup under study: a $\mathcal{PT}$-symmetric non-Hermitian superconductor of width $d$ is placed between two conventional superconductors with superconducting phases $\varphi_1$ and $\varphi_2$, respectively.}
		\label{fig:setup}
	\end{center}
\end{figure}

In this work, we study Andreev bound states in a junction between a $\mathcal{PT}$-symmetric non-Hermitian superconductor (PTS) and two conventional superconductors with phases $\varphi_1$ and $\varphi_2$ \cite{footnote}, see Fig. \ref{fig:setup}. We find that Andreev bound states in this S-PTS-S junction exist only for certain discrete values of $\varphi_1$ and $\varphi_2$ in contrast to conventional SNS (superconductor-normal-superconductor) junctions, where the spectrum is continuous with respect to the phase difference $\varphi_1-\varphi_2$. This striking difference can be attributed to the absence of Hermiticity in PTS and the combination of symmetries in the S-PTS-S junction. Remarkably, we discover decaying and growing (with respect to time) Andreev bound states at phases $0$ and $\pi$. The growth and decay occurs due to non-Hermitian electron-electron interaction in the PTS. For the phases $\pi/2$ and $3\pi/2$, we find a Majorana zero mode that moves through the PTS in the direction of the superconductor with the phase $\pi/2$. This implies that there is a supercurrent between superconductors at the phase difference $\pi$, which reminds us of a $\pi$-junction in Hermitian systems.

{\it Model.--} We consider a system of three 1D superconducting leads: two conventional superconductors and a PTS of width $d$ between them, see Fig. \ref{fig:setup}. The experimental realization of it can, for instance, be a superconducting nanowire in presence of external driving in the center region in order to induce non-Hermitian superconductivity. This nanowire shall then be attached to two loops threaded by magnetic fluxes in order to vary the phase differences between the conventional superconductors and the PTS. 

Note that parity (or space inversion) symmetry $\mathcal{P}$ inverts the sign of momentum and time-inversion symmetry $\mathcal{T}$ inverts the signs of momentum, spin, and imaginary $i$. As a result, $\mathcal{PT}$-symmetric Hamiltonians of a single-band superconductor must obey the relation $H_{PT}=H_{PT}^*$, i.e. all mean fields must be real. In order to obtain a non-Hermitian superconductor, we take the particle mean field with the opposite sign to the hole mean field, $\Delta_{PT}=-\bar{\Delta}_{PT}$ \cite{kornich:prr22}. This choice implies that the electron-electron interaction $U=\sum_{p,q}\psi^\dagger_{p+q,\uparrow}\psi^\dagger_{-p-q,\downarrow} V(q)\psi_{-p,\downarrow}\psi_{p,\uparrow}$, where $\psi$ are electron field operators, is non-Hermitian and has an asymmetric interaction potential $V(q)=-V(-q)$, meaning that when electrons interact attractively the corresponding holes interact repulsively and vice versa \cite{kornich:prr22}. This can happen due to external influence on the system, e.g. in Ref. \cite{kornich:prr22} we propose how to induce it via spatiotemporal modulations. In this case, the spatiotemporal modulations induce an asymmetry of the phonon spectrum, leading to asymmetric electron-electron interaction mediated by the phonons.

\begin{figure}[tb]
	\begin{center}
		\includegraphics[width=\linewidth]{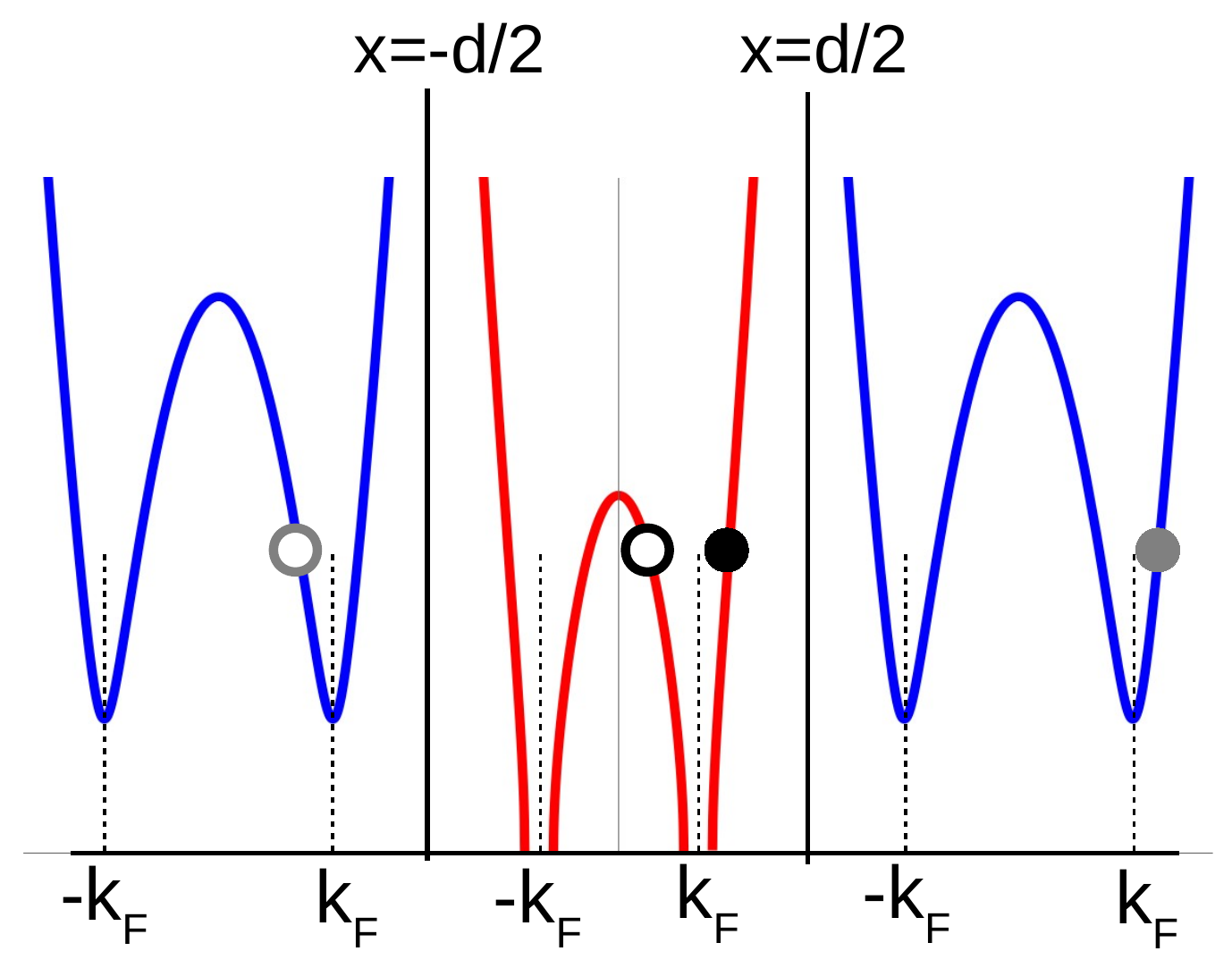}
		\caption{Scattering of the quasiparticles of PTS around $k_F$ at the boundaries with conventional superconductors. The right-moving quasiparticle (black filled circle) scatters into the right-moving quasiparticle in a conventional superconductor (grey filled circle) and a left-moving quasiparticle in PTS (black empty circle). The left-moving quasiparticle in PTS scatters at $x=-d/2$ back into the right-moving quasiparticle in PTS and into a left-moving quasiparticle in a conventional superconductor (grey empty circle). Here, we show the case, when the spectrum of PTS is real with constant $\Delta_{PT}=\Delta$ and the excitation energy of quasiparticles is chosen above the gap.  }
		\label{fig:scattering}
	\end{center}
\end{figure}

We assume that all leads form good contacts at the interfaces and there is no normal scattering there. In order to consider Andreev scattering in this junction, we represent the electron field operators for all superconductors as quasiparticles close to Fermi wavevectors $k_F$ and $-k_F$: $(\Psi^{\alpha}_{\sigma}(x))^\dagger=(\Psi^{\alpha}_{+,\sigma}(x))^\dagger e^{ik_Fx}+(\Psi^{\alpha}_{-,\sigma}(x))^\dagger e^{-ik_Fx}$, where $\alpha$ is an index denoting the type of superconductor, $\sigma$ denotes spin, and $x$ is the coordinate along the junction. In the bases $(\Psi^\alpha_l)^\dagger=\{(\Psi^\alpha_{+,\uparrow}(k))^\dagger,\Psi^\alpha_{-,\downarrow}(-k),(\Psi^\alpha_{-,\downarrow}(k))^\dagger, \Psi^\alpha_{+,\uparrow}(-k)\}$, the linearized Hamiltonians for PTS and conventional superconductors are 
\begin{eqnarray}
H_{PT}&=&v_F k\tau_z\otimes\sigma_z+\Delta_{PT} i\tau_0\otimes\sigma_y,\\
H_{1,2}&=&v_F k\tau_z\otimes\sigma_z+\Delta\tau_0\otimes(e^{i\varphi_{1,2}}\sigma_++e^{-i\varphi_{1,2}} \sigma_-),\ \ \ 
\end{eqnarray}
respectively. Here, $v_F$ is the Fermi velocity and $\otimes$ denotes the Kronecker product. In this work, we assume for simplicity that all three leads have the same Fermi velocity and Fermi wavevector. As we consider spin-singlet superconductivity and assume for simplicity that $\Delta_{PT}$ is even in frequency, then $\Delta_{PT}$ must be odd in momentum due to the fact that $V(q)$ is odd \cite{kornich:prr22}. This is in contrast to the conventional case, where the interaction potential is even and even-frequency spin-singlet pairing requires even parity in order to obey Fermi statistics. Taking into account that we consider scattering around $\pm k_F$ and $|k|\ll k_F$, we approximate the mean fields as $\Delta_{PT}(k\pm k_F)\approx\Delta_{PT}(\pm k_F)=\pm \Delta_{PT}$.  

In order to obtain eigenvectors of the Hamiltonians for a given excitation energy, we first find the wavevectors $k$ via the equations $\det[H_{\alpha}-E]=0$. For PTS, $k_{PT}=\sqrt{E^2+\Delta_{PT}^2}/v_F$ and, for conventional superconductors, we have $k_c= \sqrt{E^2-\Delta^2}/v_F$. Note that $k$ denotes a small deviation of the wavevector of the quasiparticle from $\pm k_F$ (see Fig. \ref{fig:scattering}).

The eigenvectors of the Hamiltonians yield the basis for wavefunctions (we omit the spin index for brevity, as there are no spin interactions in this work):
\begin{eqnarray}
\label{eq:PhiPTplus}
\begin{pmatrix}
\Phi^{e,+}_{PT}\\ \Phi^{h,-}_{PT}
\end{pmatrix}_{R,L}&=&\begin{pmatrix}\frac{\mp\sqrt{E^2+\Delta_{PT}^2}-E}{\Delta_{PT}}\\ 1\end{pmatrix}e^{\pm ik_{PT}x},\\
\label{eq:PhiPTminus}
\begin{pmatrix}
\Phi^{e,-}_{PT}\\ \Phi^{h,+}_{PT}
\end{pmatrix}_{R,L}&=&\begin{pmatrix}\frac{\pm\sqrt{E^2+\Delta_{PT}^2}-E}{\Delta_{PT}}\\ 1\end{pmatrix}e^{\pm ik_{PT}x},\\
\label{eq:Phi12plus}
\begin{pmatrix}
\Phi^{e,+}_{1,2}\\ \Phi^{h,-}_{1,2}
\end{pmatrix}_{R,L}&=&\begin{pmatrix}\frac{\pm\sqrt{E^2-\Delta_{1,2}^2}+E}{\Delta_{1,2}}e^{i\varphi_{1,2}}\\ 1\end{pmatrix}e^{\pm ik_cx},\\
\label{eq:Phi12minus}
\begin{pmatrix}
\Phi^{e,-}_{1,2}\\ \Phi^{h,+}_{1,2}
\end{pmatrix}_{R,L}&=&\begin{pmatrix}\frac{\mp\sqrt{E^2-\Delta_{1,2}^2}+E}{\Delta_{1,2}}e^{i\varphi_{1,2}}\\ 1\end{pmatrix}e^{\pm ik_cx},
\end{eqnarray}
Here, indices $L$ and $R$ denote left- (at $-k_c$ and $-k_{PT}$) and right-moving (at $k_c$ and $k_{PT}$) quasiparticles, respectively. Eqs. (\ref{eq:PhiPTplus}) and (\ref{eq:Phi12plus}) describe quasiparticles near $k_F$, while Eqs. (\ref{eq:PhiPTminus}) and (\ref{eq:Phi12minus}) describe quasiparticles near $-k_F$.

{\it Spectrum.--}
Assuming that the interfaces between the superconductors are ideal, we write the continuity equations for the wavefunctions. We consider Andreev scattering: right-moving quasiparticles in PTS are reflected into left-moving quasiparticles. They penetrate into the conventional superconductors as evanescent modes, if their excitation energy $E$ is smaller than $\Delta$. This gives two equations with unknown coefficients $\alpha$, $\beta$, and two transmission coefficients $t_1$ and $t_2$ at the boundaries $x=-d/2$ and $x=d/2$, respectively, for scattering of quasiparticles close to $k_F$ and close to $-k_F$. For an illustration of scattering processes in an S-PTS-S junction with excitation energies of quasiparticles above the gap, see Fig. \ref{fig:scattering}. The equations for scattering of quasiparticles near $k_F$ are
\begin{eqnarray}
\label{eq:continuityeq}
\nonumber \left(\alpha\begin{pmatrix}
\Phi^{e,+}_{PT}\\ \Phi^{h,-}_{PT}
\end{pmatrix}_R+\beta\begin{pmatrix}
\Phi^{e,+}_{PT}\\ \Phi^{h,-}_{PT}
\end{pmatrix}_L\right)\Bigg|_{x=\mp d/2}=\\ =t_{1,2}\begin{pmatrix}
\Phi^{e,+}_{1,2}\\ \Phi^{h,-}_{1,2}
\end{pmatrix}_{L,R}\Bigg|_{x=\mp d/2}.
\end{eqnarray}
From this equations, we obtain the spectrum of Andreev bound states determined by

\begin{eqnarray}
\label{eq:reqr}
\nonumber
\frac{e^{-ikd}(k_{PT}v_F\Delta-e^{i\varphi_1}\Delta_{PT}k_cv_F+E(e^{i\varphi_1}\Delta_{PT}+\Delta))}{k_{PT}v_F\Delta+e^{i\varphi_1}\Delta_{PT}k_cv_F-E(e^{i\varphi_1}\Delta_{PT}+\Delta)}=\\ =\frac{e^{ikd}(k_{PT}v_F\Delta+e^{i\varphi_2}\Delta_{PT}k_cv_F+E(e^{i\varphi_2}\Delta_{PT}+\Delta))}{k_{PT}v_F\Delta-e^{i\varphi_2}\Delta_{PT}k_cv_F-E(e^{i\varphi_2}\Delta_{PT}-\Delta)}. \ \ \ \ \
\end{eqnarray}

The equations for scattering of quasiparticles near $-k_F$ have the same form as Eq.~(\ref{eq:continuityeq}), but with the corresponding wavefunctions from Eqs.~(\ref{eq:PhiPTminus}) and (\ref{eq:Phi12minus}). 

Taking into account that we consider a composite non-Hermitian system, the energy of Andreev bound states can be complex. One way to solve Eq.~(\ref{eq:reqr}) is to divide it into real and imaginary parts and find common roots for real and imaginary parts of energy. Remarkably, in contrast to conventional SNS junctions, there are only certain discrete values of phases $\varphi_1$ and $\varphi_2$ with valid solutions. We have found that pairs $\{0,\pi\}$ and $\{\pi/2,3\pi/2\}$ give solutions. 

{\it Phases $0$ and $\pi$.--} For the short junction, $\Delta d/v_F\ll 1$, and $\varphi_1=0$ and $\varphi_2=\pi$, we obtain $E=i\Delta_{PT}\Delta d/v_F$ for scattering close to $k_F$ and the complex conjugate of it, $E=-i\Delta_{PT}\Delta d/v_F$, for scattering close to $-k_F$. Once we reverse the phases, $\varphi_1=\pi$ and $\varphi_2=0$, we obtain the opposite results: $E=i\Delta_{PT}\Delta d/v_F$ for scattering close to $-k_F$ and $E=-i\Delta_{PT}\Delta d/v_F$ for scattering close to $k_F$. This means that the phases define, which scattering process dominates, because negative imaginary energy implies decay and positive imaginary energy implies growth or pumping of the state with respect to time. 

We can draw an analogy to Andreev bound states in conventional SNS junctions. At the phase difference $\pi$ and transparent interfaces, the processes of scattering around $k_F$ and $-k_F$ yield degenerate states at $E=0$. However, if $0<\varphi_2-\varphi_1<\pi$, the states for $E>0$ are formed by the scattering at $-k_F$ and if $\pi<\varphi_2-\varphi_1<2\pi$, the states for $E>0$ are formed due to scattering at $k_F$. For $E<0$, it is vice versa. Thus, in conventional SNS junctions, the phases also define which process dominates at a certain energy. The striking difference to our junction is that in the S-PTS-S case, the degeneracy at phase difference $\pi$ is lifted because we obtain non-stationary states that decay or grow. Moreover, in contrast to conventional SNS junctions, the electron and hole content of the Andreev bound states at the edges of the junction are not equal. 

Let us discuss in more detail, what growth and decay of these bound states mean physically in our system. The process of growth and decay occurs within the PTS. If we look at a very narrow layer of the PTS, $d\rightarrow 0$, we can see that $E\rightarrow 0$, corresponding to the value of energy at the phase difference $\pi$ in clean conventional SNS junctions without normal scattering. This happens because at $d\rightarrow 0$ electrons penetrate through the PTS region quickly and do not experience the influence of its properties. Formation of an Andreev bound state at energy $\varepsilon_n$ in conventional SNS junction can be understood as an interference effect, when the phase difference between electron and hole in one scattering cycle between the interfaces of superconductors match the phase difference of superconductors plus $2\arccos{(\varepsilon_n/\Delta)}+2\pi n$ \cite{sellier:prb03}. In our case, it can also be understood as interference, but of decaying and growing wavefunctions. The decaying and growing states appear in PTS in the $k$ gaps, see Fig.~\ref{fig:scattering}, where the eigen energies $\pm\sqrt{(k^2/(2m)-\mu)^2-\Delta^2}$ (before linearization) are imaginary. Microscopically, we understand the processes of decay and growth of the state as the decay and formation of quasiparticles in the PTS due to non-Hermiticity of electron-electron interaction, i.e. (anti-)correlation of electrons and holes. This happens because they interact attractively or repulsively depending on the direction of the momentum transferred during electron-electron interaction \cite{kornich:prr22}. Note that the spectrum of the PTS is gapless. Hence, it costs no energy to add/remove quasiparticles to/from it.

{\it Perturbation theory.--} We now verify that the allowed phases are restricted to discrete values via perturbation theory in $\Delta_{PT}$. We assume that $\Delta_{PT}$ is the smallest energy scale in our system. This means that we should exclude $E=0$. 

Let us consider the case of scattering at $-k_F$ for concreteness. We note that the upper term in the left-mover eigenvector Eq. (\ref{eq:PhiPTminus}) diverges in the limit $\Delta_{PT}\rightarrow 0$. This happens because we have put all normalizations into the factors $\alpha$, $\beta$, and $t_{1,2}$ in Eq. (\ref{eq:continuityeq}). Consequently, we obtain $\lim_{\Delta_{PT}\rightarrow 0}\beta=0$. Therefore in order to analyze this limit, we need to multiply $\beta$ by $\Phi_{PT}^{e,-}$, expand in $\Delta_{PT}$, and then equalize the factors obtained from the equations describing scattering at $x=-d/2$ and at $x=d/2$ (analogous to Eqs. (\ref{eq:continuityeq}), but for scattering close to $-k_F$). Retaining only linear terms in $\Delta_{PT}$ and assuming short junctions, $Ed/v_F\ll 1$, we express $\Delta_{PT}$ in terms of $E$ and the phases $\varphi_1$ and $\varphi_2$:
\begin{eqnarray}
\label{eq:DeltaPT}
\Delta_{PT}=\frac{iE(1-e^{i(\varphi_2-\varphi_1)})-\sqrt{\Delta^2-E^2}(e^{i(\varphi_2-\varphi_1)}+1)}{v_F^{-1}\Delta d(e^{-i\varphi_1}-e^{i\varphi_2})}.\ \ \ \ \ 
\end{eqnarray}
We know that in order to have a PTS, $\Delta_{PT}$ must be real. This can happen for $|E|<\Delta$, if $E$ is purely imaginary, $E=iE'$ and $E'\in \Re$, and $\varphi_2-\varphi_1=(2n+1)\pi$, $\varphi_1=m\pi$, and $\varphi_2=l \pi$, where $n$, $m$, and $l$ are integer numbers. 

Thus, we substitute $E=iE'$ into Eq. (\ref{eq:DeltaPT}). We also assume that $E'\ll \Delta$, as we treat the PTS perturbatively. We obtain
\begin{eqnarray}
E'=i\Delta\cot{\left(\frac{\varphi_1-\varphi_2}{2}\right)}+\Delta_{PT}\Delta\frac{ d}{v_F}\frac{\sin{\left(\frac{\varphi_1+\varphi_2}{2}\right)}}{\sin{\left(\frac{\varphi_1-\varphi_2}{2}\right)}}.\ \ \ \ \
\end{eqnarray}
We can see that in order for $E'$ to be real and finite, we need $\varphi_1-\varphi_2=(2n+1)\pi$. If we exchange the phases $\varphi_1$ and $\varphi_2$, the sign of $E'$ changes. For instance, we obtain $E'<0$ for $\varphi_1=0$ and $\varphi_2=\pi$, as stated in the previous section.

{\it Phases $\pi/2$ and $3\pi/2$.--} For these phases, we obtain $E=0$ from the full calculation using Eq. (\ref{eq:reqr}) and an analogous equation for scattering around $-k_F$. Let us first discuss, why there is no pumping or decay in this case. 

For $\varphi_1=3\pi/2$ and $\varphi_2=\pi/2$, we obtain
\begin{eqnarray}
\beta&=&0,\\
\label{eq:t1}
t_1&=&e^{(\Delta-i\Delta_{PT})d/(2v_F)},\\
\label{eq:t2}
t_2&=&e^{(\Delta+i\Delta_{PT})d/(2v_F)}.
\end{eqnarray} 
This implies that the eigenvectors in all three superconductors are the same with the exception of the factors $e^{\pm ik_{PT}x}$, $e^{\pm ik_{c}x}$; meaning that if we put $x=0$, they are equal. For the scattering close to $k_F$ and putting $x=0$, they are
\begin{eqnarray}
\begin{pmatrix}
\Phi^{e,+}_{PT}\\ \Phi^{h,-}_{PT}
\end{pmatrix}_{R}=\begin{pmatrix}
\Phi^{e,+}_{1,2}\\ \Phi^{h,-}_{1,2}
\end{pmatrix}_{L}=\begin{pmatrix}
\Phi^{e,+}_{1,2}\\ \Phi^{h,-}_{1,2}
\end{pmatrix}_{R}=\begin{pmatrix}-1\\ 1\end{pmatrix}.
\end{eqnarray}
For the subspace close to $-k_F$, the corresponding eigenvectors are $(1\ \ 1)^T$. This means that there is only a right-moving quasiparticle in PTS. It propagates through PTS obtaining a phase $\Delta_{PT} d/v_F$. This quasiparticle penetrates through the whole junction without reflection because the eigenmode is the same in all three superconductors. It becomes an evanescent mode in the conventional superconductors, as its energy is zero, i.e. less than $\Delta$.

In the inverse case, $\varphi_1=\pi/2$ and $\varphi_2=3\pi/2$, the quasiparticle is left-moving, because $\alpha$ vanishes in Eq. (\ref{eq:continuityeq}). This implies that there is only a left-moving quasiparticle in the PTS part of the S-PTS-S junction for these phases. The general effect is analogous to the previous case: A quasiparticle obtains a phase $-\Delta_{PT}d/v_F$ within the PTS. It does not scatter at the interfaces and becomes an evanescent mode within the conventional superconductors. Its eigenvector is $(1\ \ 1)^T$ in the subspace close to $k_F$ and is $(-1\ \ 1)^T$ for the subspace close to $-k_F$.

This non-reciprocal behaviour can be understood as a supercurrent of quasiparticles. In conventional SNS junctions, the supercurrent often times follows the sinusoidal current-phase relation $I\propto \sin{(\varphi_1-\varphi_2)}$. Hence, the phase difference $\pi$ gives zero current. However, in our S-PTS-S junction, there is a supercurrent of quasiparticles with net zero charge. The direction of the current is defined by the phases: It moves in the direction of the superconductor with phase $\pi/2$. As these quasiparticles are at zero energy and all superconductors obey particle-hole symmetry (see section ``Symmetry considerations''), these states are Majorana zero modes according to the definition for non-Hermitian Hamiltonians \cite{kawabata:prx19}.

{\it Symmetry considerations.--} We now argue why the phases $\{0,\pi\}$ and $\{\pi/2,3\pi/2\}$ allow for solutions and why they are physically different in terms of symmetry considerations. 

In conventional superconductors, quasiparticles have particle-hole symmetry defined as $\mathcal{C}_cH_{1,2}^*(k)\mathcal{C}_c^{-1}=\mathcal{C}_cH_{1,2}^T(k)\mathcal{C}_c^{-1}=-H_{1,2}(-k)$, where $\mathcal{C}_c$ is a unitary transformation and the subscript $T$ denotes transposition. Particle-hole symmetry of non-Hermitian systems has two definitions due to the fact that $H^*\neq H^T$ \cite{kawabata:prx19}. PTS obeys the particle-hole symmetry defined as $\mathcal{C} H_{PT}^T(k)\mathcal{C}^{-1}=-H_{PT}(-k)$, where $\mathcal{C}$ is a unitary transformation. Both types of particle-hole symmetries require pairs of eigenvalues $\{E,-E\}$, while the first one also requires either $\Re\{E\}=0$ or pairs $\{E,-E^*\}$. This means that in order to satisfy all symmetries of the constituents of the junction, the energies must be either purely real or purely imaginary and must appear with both signs in the spectrum.

In the case of phases $\{0,\pi\}$, we obtain a purely imaginary spectrum. When we have phases $\{\pi/2,3\pi/2\}$, the energy is zero. In principle, obtaining a real energy spectrum in a non-Hermitian system is non-trivial. It was considered to be not feasible for a long time, until it was shown by Bender and Boettcher for $\mathcal{PT}$-symmetric systems in 1998 \cite{bender:prl98}. Let us consider the effect of $\mathcal{PT}$ symmetry on S-PTS-S junction. The conventional superconductors are transformed as $\mathcal{PT}H_{1,2}(\mathcal{PT})^{-1}=H_{1,2}^*$. Taking into account that we have phases $\pi/2$ and $3\pi/2$, $H_1$ and $H_2$ exchange with each other under this transformation. However, $\mathcal{P}$ symmetry is by definition $x\rightarrow -x$, which means that if we act on the whole S-PTS-S junction with $\mathcal{PT}$ symmetry, we have to exchange the conventional superconductors. Thus, $\mathcal{P}$ and $\mathcal{T}$ effectively exchange the conventional superconductors twice and we obtain the same junction. This means that S-PTS-S junctions with phases $\pi/2$ and $3\pi/2$ are $\mathcal{PT}$-symmetric as a whole, and real energy of the Andreev bound state can be expected. 

{\it Conclusions.--} In this work, we consider a junction between two conventional superconductors and a $\mathcal{PT}$-symmetric non-Hermitian superconductor. $\mathcal{PT}$-symmetric non-Hermitian superconductivity can emerge from Dzyaloshinskii-Moriya interaction in combination with an external bath or the imbalance between electron-electron and hole-hole pairs \cite{ghatak:prb18} or via spatiotemporal modulation of a material with two interacting phonon bands \cite{kornich:prr22}.

We have shown that Andreev bound states are formed in the S-PTS-S junction. However, the physical properties of these states are conceptually different from the ones that occur in ordinary SNS junctions. First of all, solutions exist only for particular discrete values of phases of the conventional superconductors. Such strong selection can be useful in logical quantum devices. Remarkably, for phases $0$ and $\pi$, we obtain growing and decaying states. These non-equilibrium processes occur due to non-Hermitian electron-electron interaction in PTS. Experimentally, this can lead to the presence of only one strongly-pumped mode. This pumped mode indicates particular phase differences by growing signal in time. In this sense, it corresponds to a novel feature of a Josephson junction as an amplifier. For phases $\pi/2$ and $3\pi/2$, there is a unidirectional supercurrent of Majorana zero modes. The discreteness of $\varphi_1$ and $\varphi_2$ is stable with respect to a change of the length of the PTS and for moderate temperatures (below superconducting transition temperatures). 
\begin{acknowledgments}
	We acknowledge useful discussions with Sebastian Bergeret. This work was supported by the DFG (SPP1666 and SFB1170 ``ToCoTronics''), the W{\"u}rzburg-Dresden Cluster
	of Excellence ct.qmat, EXC2147, project-id 390858490, and the Elitenetzwerk Bayern Graduate School on ``Topological Insulators''. We thank the Bavarian Ministry of Economic Affairs, Regional Development and Energy for financial support within the High-Tech Agenda Project ``Bausteine f\"ur das Quanten Computing auf Basis topologischer Materialen''.
\end{acknowledgments}


\begin{thebibliography}{}
\bibitem{footnote}{Physically, the effect of superconductors with phases $\varphi_1$ and $\varphi_2$ occurs due to contact with the PTS. Hence, we can understand the dependence on $\varphi_1$ and $\varphi_2$ as the dependence on the phase differences between superconductor 1 and PTS and superconductor 2 and PTS.}
\end{thebibliography}
\end{document}